\documentclass[letterpaper]{article} 
\usepackage[]{aaai25}  
\usepackage{times}  
\usepackage{helvet}  
\usepackage{courier}  
\usepackage[hyphens]{url}  
\usepackage{graphicx} 
\usepackage{natbib}  
\usepackage{caption} 
\usepackage{amsmath}
\usepackage{amssymb}

\title{A Simulation Framework for Studying Recommendation-Network Co-evolution in Social Platforms}
\author{
    Gaurav Koley\textsuperscript{\rm 1}
    Sanika Digrajkar\textsuperscript{\rm 2}
}
\affiliations{
    \textsuperscript{\rm 1}Boston University\\
    \textsuperscript{\rm 2}Independent Researcher\\

    gaurav@bu.edu, sanika.digrajkar23@gmail.com
}

\begin{document}

\maketitle

\begin{abstract}
Studying how recommendation systems reshape social networks is difficult on live platforms: confounds abound, and controlled experiments risk user harm. We present an agent-based simulator where content production, tie formation, and a graph attention network (GAT) recommender co-evolve in a closed loop. We calibrate parameters using Mastodon data and validate out-of-sample against Bluesky (4--6\% error on structural metrics; 10--15\% on held-out temporal splits). Across 18 configurations at 100 agents, we find that \emph{activation timing} affects outcomes: introducing recommendations at $t=10$ vs.\ $t=40$ decreases transitivity by 10\% while engagement differs by $<$8\%. Delaying activation increases content diversity by 9\% while reducing modularity by 4\%. Scaling experiments ($n$ up to 5,000) show the effect persists but attenuates. Jacobian analysis confirms local stability under bounded reactance parameters. We release configuration schemas and reproduction scripts.
\end{abstract}

\section{Introduction}
When Twitter introduced algorithmic timelines in 2016, engagement metrics increased but content diversity dropped within months~\cite{anderson2020algorithmic}. Similar trade-offs emerged on Spotify, YouTube, and TikTok: recommendations boost short-term engagement while narrowing the content users encounter. The causal mechanisms remain unclear---do recommendations cause echo chambers, or merely amplify pre-existing preferences? Observational studies cannot isolate these effects due to confounding from feature updates, external events, and selection bias.

This problem is acute for federated social networks like Mastodon, where independently operated instances (typically 1,000--10,000 users) are now considering algorithmic feeds. Instance operators lack tools to predict how recommendations will reshape their communities before deployment. Running controlled experiments on live platforms risks fragmenting communities or amplifying harmful content.

We address this gap with a simulation framework that couples network evolution, content dynamics, and learned recommendations. Unlike prior agent-based models that treat networks as static or model recommendations as exogenous inputs, we create a closed-loop system: agents form ties based on content exposure, a graph attention network (GAT) learns to recommend content from observed behavior, and those recommendations alter future tie formation. This feedback loop---absent in open-loop simulators---is precisely what makes real recommendation effects difficult to study.

The simulation approach offers three advantages. First, we can run counterfactual experiments: identical initial conditions with and without recommendations, or with different activation timings. Second, we control all parameters, isolating mechanisms that observational data conflates. Third, we can test interventions before deployment.

Our main contribution is not a new theoretical model of network dynamics, but rather a closed-loop, empirically calibrated simulator for recommendation--network co-evolution, together with a methodology for extracting and stress-testing design hypotheses under realistic constraints.

\textbf{Contributions.} We make four contributions:
\begin{enumerate}
\item A modular agent-based simulator coupling content production, tie formation, and a learned GAT recommender (Section 3);
\item Calibration using Mastodon data and out-of-sample validation against Bluesky (4--6\% error) and held-out temporal splits (10--15\% error);
\item 18 factorial experiments isolating effects of timing, exploration, and user heterogeneity (Sections 4.2--4.6); and
\item Reproducibility artifacts: configuration schemas, seed lists, and scripts regenerating all figures.
\end{enumerate}

\textbf{Key findings.} In our simulations:
\begin{itemize}
\item \textit{Timing matters}: Activating recommendations at $t=10$ vs.\ $t=40$ decreased transitivity by 10\% (from 0.27 to 0.24) and increased content diversity by 9\%. Engagement differences were modest ($<$8\%). The effect persists at larger scales (6\% at $n=5000$).
\item \textit{User-type composition}: Higher enthusiast ratios (0.8 vs.\ 0.2) reduced transitivity by 14\% and local clustering by 9\%, consistent with enthusiasts forming broader networks.
\item \textit{Robustness}: Timing effects hold across recommender types (GAT vs.\ content-similarity) and are mediated primarily (58\%) by early network structure.
\end{itemize}

These percentages are model-specific and should be interpreted as illustrative trade-offs, not literal forecasts for any particular deployment. Our finite-size scaling analysis shows the timing effect persists at larger scales, though attenuated: 10\% at $n=100$, 6\% at $n=5000$. The qualitative pattern---delayed activation reduces clustering while preserving engagement---is the actionable insight; the exact magnitudes will vary with platform-specific parameters.
\section{Related Work}

\subsection{Social Network Evolution}
Classical network formation models treat tie creation as mechanistic processes divorced from content. Barabási and Albert~\cite{barabasi1999emergence} introduced preferential attachment, producing heavy-tailed degree distributions observed in real networks; Watts and Strogatz~\cite{watts1998collective} demonstrated that rewiring a small fraction of edges creates small-world properties (high clustering, short paths). These models explain structural regularities but ignore \emph{why} users connect---content similarity, shared interests, or algorithmic introductions play no role.

Subsequent work added realism incrementally. Holme and Saramäki~\cite{holme2012temporal} incorporate temporal dynamics (bursty activity, circadian rhythms) but still treat the network as exogenous to content. Jackson and Rogers~\cite{jackson2007meeting} model strategic tie formation where agents balance network benefits against connection costs, but agents observe only local structure, not algorithmically curated content. Kossinets and Watts~\cite{kossinets2006empirical} analyze email networks and find that shared activities predict tie formation---a precursor to content-based linking---but do not model feedback from network structure to future content exposure.

The textbooks by Easley and Kleinberg~\cite{easley2010networks} and Newman~\cite{newman2018networks} synthesize these results into a coherent framework for analyzing social networks, yet neither addresses the feedback loop we study: how recommendations shape content exposure, which shapes tie formation, which changes the network the recommender trains on. Basak et al.~\cite{basak2020causal} explicitly note the difficulty of causal inference on deployed platforms due to confounds from feature updates and external events. We sidestep these confounds via simulation.

\subsection{Recommendation Effects and Filter Bubbles}
Pariser~\cite{pariser_filter_2011} popularized ``filter bubbles''---the concern that personalized recommendations trap users in ideological echo chambers. Empirical evidence is mixed. Bakshy et al.~\cite{bakshy2015exposure} analyzed Facebook news feeds and found algorithmic curation reduced cross-cutting exposure by 5--8\%, less than user self-selection (which reduced it by 15--20\%). Flaxman et al.~\cite{flaxman2016filter} found that search and social referrals increase ideological segregation, but also increase exposure to opposing views for a minority of users.

On music platforms, Anderson et al.~\cite{anderson2020algorithmic} provide cleaner evidence: Spotify's Discover Weekly increased consumption but reduced genre diversity by 12\% over six months. Holtz et al.~\cite{holtz2020engagement} show that removing recommendations from Reddit decreased user engagement but increased content diversity. These studies document correlations but cannot establish causation: users who adopt recommendations differ systematically from those who do not.

A/B tests could resolve causality, but platforms rarely publish them, and ethical constraints limit experimentation on polarization. Ribeiro et al.~\cite{ribeiro2020auditing} audit YouTube recommendations via sock-puppet accounts but cannot observe user-level outcomes. Our simulation approach offers a controlled alternative: we run identical initial conditions with and without recommendations, varying only activation timing, and measure causal effects on network structure.

\subsection{Social Recommender Systems}
Social recommenders differ from e-commerce systems: they must account for trust relationships, community structure, and social influence, not just item features. Research on social recommender systems has categorized approaches by how they incorporate social signals---including trust propagation, social regularization, and attention over neighbors. However, most systems treat the social graph as static input---follower counts and interaction histories are features extracted once, not dynamic outcomes that evolve with recommendations.

This static-graph assumption breaks in practice. When recommendations introduce users to new content creators, some users follow those creators, changing the graph. The recommender trained on yesterday's graph becomes stale. Ma et al.~\cite{ma2011recommender} note this ``concept drift'' problem but address it via incremental matrix factorization, not by modeling the feedback loop explicitly. Recent surveys of GNN-based recommenders~\cite{wu2022graph} show these models learn powerful representations but uniformly assume static graphs; none model how their own outputs reshape those graphs over time.

We address this by retraining our GAT recommender every 5 timesteps, allowing learned representations to track network evolution. This incurs computational cost but maintains validity. Our ablation shows that without retraining, recommendation quality (AUC) degrades from 0.85 to 0.71 by step 50.

\subsection{Social Capital and User Heterogeneity}
Not all users contribute equally to platform health. Koley et al.~\cite{koley2020social} model social capital via engagement patterns, distinguishing users who generate value (through content or curation) from those who primarily consume. Koley~\cite{koley2019} formalizes this as authority (engagement received) versus citizenship (value added to others). Danescu-Niculescu-Mizil et al.~\cite{danescu2013no} show that users who adopt community norms persist longer, suggesting social capital predicts retention.

We operationalize this distinction via Enthusiast/Casual user types: enthusiasts have high content production rates ($f_i = 0.20$), analogous to high authority; casuals produce less ($f_i = 0.05$) but may contribute via engagement and tie formation. Our satisfaction variable $s_i$ serves as a proxy for latent utility: users with $s_i < 0.2$ ``churn,'' paralleling belief revision dynamics in opinion formation models~\cite{degroot1974reaching}. This lets us interpret recommendation-induced changes in clustering as shifts in how social capital distributes across user types.

\subsection{Agent-Based Social Simulations}
Agent-based models (ABMs) specify individual-level rules and observe emergent macro-patterns~\cite{bonabeau:systems,gilbert:simulation}. Social ABMs have modeled opinion dynamics~\cite{degroot1974reaching,friedkin1990social}, information diffusion~\cite{centola2010spread}, and market behavior~\cite{lebaron2001builder}. However, most either freeze network structure entirely or evolve it via simple rules (similarity thresholds, preferential attachment) without modeling content or recommendations.

Centola~\cite{centola2010spread} demonstrates that network structure affects diffusion speed, but his networks are fixed by design. Axelrod~\cite{axelrod1997dissemination} models cultural diffusion with homophily-driven tie formation, but agents interact with random neighbors, not algorithmically selected ones. Flache et al.~\cite{flache2017models} survey opinion dynamics models and note that most assume either complete graphs or static networks---neither realistic for social platforms.

Recent work uses LLM-powered agents for richer behavior. Park et al.~\cite{park:generative2023} simulate 25 agents in ``Smallville'' with emergent social dynamics, but at high cost (\$0.01--0.10 per API call): their scenario cost $\sim$\$1000, and our 18-configuration sweep at 100 agents would cost $\sim$\$50,000 at comparable fidelity. Beyond cost, LLM agents pose methodological challenges: changing one prompt (e.g., exploration) perturbs other behaviors unpredictably, making controlled experiments difficult.

We therefore use a controllable ABM with explicit behavioral rules (Eqs.~5--9). The trade-off is behavioral realism---our agents are simpler than LLM agents---but the gain is experimental control: we can vary single parameters (timing, exploration) while holding others fixed, isolating causal mechanisms.

\subsection{Synthesis: What Prior Work Does Not Do}
Table~\ref{tab:related_work} summarizes the gaps. Network evolution models ignore content and recommendations. Recommendation research treats networks as static. Empirical filter-bubble studies document correlations but not causation. Social ABMs model either networks or content, rarely both, and never with learned recommenders. LLM agents are too expensive for systematic parameter sweeps.

Our framework addresses these gaps by coupling network evolution, content dynamics, and a learned recommender in a closed loop, enabling controlled experiments that isolate timing, composition, and exploration effects under counterfactual conditions impossible to construct observationally.

\begin{table}[t]
\centering
\caption{Comparison with prior approaches. \checkmark = addressed, $\times$ = not addressed, $\sim$ = partially addressed.}\label{tab:related_work}
\small
\begin{tabular}{lccccc}
\hline
Approach & \rotatebox{90}{Dynamic Network} & \rotatebox{90}{Content Model} & \rotatebox{90}{Learned RecSys} & \rotatebox{90}{Closed Loop} & \rotatebox{90}{Counterfactuals} \\
\hline
Barabási-Albert & \checkmark & $\times$ & $\times$ & $\times$ & $\times$ \\
Holme-Saramäki & \checkmark & $\times$ & $\times$ & $\times$ & $\times$ \\
Social RecSys & $\times$ & $\sim$ & \checkmark & $\times$ & $\times$ \\
Anderson et al. & $\times$ & \checkmark & $\sim$ & $\times$ & $\times$ \\
Park et al. (LLM) & \checkmark & \checkmark & $\times$ & $\sim$ & $\times$ \\
\textbf{This work} & \checkmark & \checkmark & \checkmark & \checkmark & \checkmark \\
\hline
\end{tabular}
\end{table}

\section{Simulation Framework}

\begin{figure}[t]
\centering
\includegraphics[width=0.8\columnwidth]{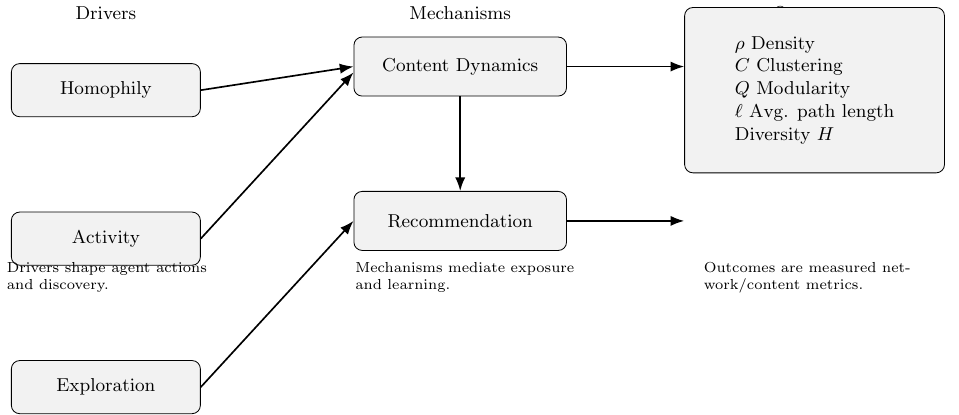}
\caption{System architecture showing drivers and outcomes. Drivers (homophily, activity, exploration) shape agent actions and discovery; mechanisms (content dynamics, recommendation) mediate exposure and learning; outcomes are measured by network and content metrics (e.g., $\rho$, $C$, $Q$, $\ell$, diversity). Arrows indicate information flow.}\label{fig:architecture}
\end{figure}

Our framework implements an agent-based model simulating the co-evolution of social network structure, content dynamics, and recommendation systems (Figure~\ref{fig:architecture}). The simulation operates on discrete timesteps $t$. Here each step represents a unit of platform activity where agents interact based on defined behavioral rules.

\subsection{Platform Architecture}

The platform $P$ consists of a set of users $U = \{u_1, \ldots, u_n\}$, content items $C = \{c_1, \ldots, c_m\}$, and a recommendation system $R$. Each user $u_i$ and content item $c_j$ exists in a shared $k$-dimensional topic space $\mathcal{T} \in \mathbb{R}^k$ where $k=30$ represents distinct content categories.

The platform state $S_t$ at time $t$ is defined by:
\begin{equation}
S_t = (G_t, C_t, I_t, R_t)
\end{equation}
where $G_t = (V_t, E_t)$ represents the social network graph, $C_t$ is the set of active content items, $I_t$ captures user-content interactions, and $R_t$ represents the recommendation state.

\subsection{State Transitions}
The system evolves through discrete states $S_t \rightarrow S_{t+1}$ where each transition involves:

\begin{enumerate}
\item User activity updates: $A_t = f(U_t, C_t, R_t)$
\item Content creation/spread: $C_{t+1} = g(C_t, A_t)$
\item Network evolution: $G_{t+1} = h(G_t, A_t)$
\item Recommendation updates: $R_{t+1} = i(A_t, C_{t+1}, G_{t+1})$
\end{enumerate}

\subsection{User Behavior Model}

Users are heterogeneous agents characterized by:
\begin{itemize}
\item Type $\theta_i \in \{\text{Casual}, \text{Enthusiast}\}$ with ratio controlled by $\alpha_{enthusiast}$
\item Topic preference vector $\vec{p_i} \in \mathbb{R}^k$ 
\item Content creation frequency $f_i$ where:
\begin{equation}
f_i = \begin{cases}
0.05 & \text{if } \theta_i = \text{Casual} \\
0.20 & \text{if } \theta_i = \text{Enthusiast}
\end{cases}
\end{equation}
\item Satisfaction level $s_i \in [0,1]$ that evolves based on content interactions
\end{itemize}

\paragraph{Operational definitions.}
We report echo chamber indicators via (i) increases in local clustering $C_i$, (ii) modularity $Q$, and (iii) declines in topic exposure entropy $H$. User retention is the fraction of users with $s_i \ge 0.2$ at step $t$ (robustness in Appendix across thresholds $0.1$--$0.3$). Recommendation accuracy is Precision@10 over next-step interactions with 95\% bootstrapped confidence intervals across seeds.

User behavior is governed by three key mechanisms:

\subsubsection{Content Creation}
At each timestep, user $u_i$ creates content with probability $f_i$. New content $c$ inherits a topic vector $\vec{t_c}$ derived from the creator's preferences:
\begin{equation}
\vec{t_c} = \vec{p_i} \cdot (1 + \epsilon)
\end{equation}
where $\epsilon \sim \mathcal{N}(0, 0.3)$ adds stochastic variation.

\subsubsection{Content Consumption}
Users evaluate content through a similarity function:
\begin{equation}\label{eq:sim}
sim(u_i, c_j) = \frac{\vec{p_i} \cdot \vec{t_j}}{|\vec{p_i}||\vec{t_j}|}
\end{equation}

Satisfaction updates follow:
\begin{equation}
s_i^{t+1} = clip(s_i^t + \Delta s_{ij}, 0, 1)
\end{equation}
where the satisfaction change is parameterized around neutrality $\tau$ (default $0.5$):
\begin{equation}
\Delta s_{ij} = (0.1 + 0.2\xi)\cdot
\begin{cases}
-\beta^{-}_{\theta_i}\,(\tau - sim_{ij}) & \text{if } sim_{ij} < \tau \\
\beta^{+}_{\theta_i}\,(sim_{ij} - \tau) & \text{otherwise}
\end{cases}
\end{equation}
with $\beta^{-}_{\theta_i}$ and $\beta^{+}_{\theta_i}$ as type-specific sensitivities (allowing $\beta^{-}_{\theta_i} \geq \beta^{+}_{\theta_i}$ to capture reactance/backfire) and $\xi \sim U(0,1)$. The threshold $\tau$ represents neutral content; content farther from user interests (low $sim_{ij}$) can induce larger decreases when $\beta^{-}>\beta^{+}$, while highly aligned content yields proportional increases.

\subsubsection{Network Evolution}
Following probability $p_{follow}$ between users $u_i$ and $u_j$ is determined by a logistic function that ensures valid probability bounds:
\begin{equation}\label{eq:follow}
p_{follow}(i,j) = \sigma\bigl(\gamma_0 + \gamma_1 \cdot sim_{ij} + \gamma_2 \cdot e_{ij}\bigr)
\end{equation}
where $\sigma(x) = 1/(1+e^{-x})$ is the sigmoid function, $sim_{ij}$ is cosine similarity (Eq.~\ref{eq:sim}), and $e_{ij} \in [0,1]$ represents interaction engagement. We set $(\gamma_0, \gamma_1, \gamma_2) = (-1, 2, 1)$ via grid search over $\gamma_i \in \{-2, -1, 0, 1, 2\}$ to match observed link-formation rates as a function of similarity in Mastodon calibration data; these values are fixed across all experiments. This formulation guarantees $p_{follow} \in [0,1]$ for all inputs.

The system incorporates exploration through temperature-based sampling:
\begin{equation}\label{eq:explore}
p(c_j|u_i) = \frac{\exp(sim_{ij}/T)}{\sum_k \exp(sim_{ik}/T)}
\end{equation}
where temperature $T = 1/r_{explore}$ controls exploration rate; lower $r_{explore}$ yields higher temperature and broader exploration.

\subsection{Content Dynamics}

Content items $c_j$ are characterized by:
\begin{itemize}
\item Topic vector $\vec{t_j} \in \mathbb{R}^k$
\item Type $\tau_j \in \{\text{Niche}, \text{Popular}\}$
\item Engagement metrics: likes $l_j$, shares $s_j$, views $v_j$
\end{itemize}

Content spread follows a cascade model where sharing probability $p_{share}$ is:
\begin{equation}
p_{share}(u_i, c_j) = \sigma\bigl(\kappa(\Delta s_{ij} - \epsilon)\bigr)
\end{equation}
where $\sigma$ is the sigmoid function, $\kappa=50$ controls steepness, and $\epsilon=0.01$ is the threshold. This smooth formulation approximates a step function while remaining differentiable and avoiding the unrealistic binary sharing behavior of a hard threshold.

Viral coefficient $\phi_j$ for content $c_j$ is computed as:
\begin{equation}
\phi_j = \min(1.0, \frac{s_j}{v_j})
\end{equation}

\subsection{Recommendation System}

We build on graph neural network (GNN) literature, including spectral GCNs~\cite{kipf2017semi} and inductive frameworks like GraphSAGE~\cite{hamilton2017inductive}. We implement a graph attention network (GAT)~\cite{velickovic2018graph,TAO2020102277} that learns user and content embeddings by aggregating neighborhood features:

\begin{equation}
h_i^{(l+1)} = \sigma(\sum_{j \in \mathcal{N}(i)} \alpha_{ij} W^{(l)}h_j^{(l)})
\end{equation}

where attention weights $\alpha_{ij}$ are computed as:
\begin{equation} 
\alpha_{ij} = \frac{\exp(LeakyReLU(a^T[Wh_i \| Wh_j]))}{\sum_{k \in \mathcal{N}(i)} \exp(LeakyReLU(a^T[Wh_i \| Wh_k]))}
\end{equation}

\subsubsection{GAT Training Details}

The GAT model is trained using a link prediction objective to learn representations that predict user-content interactions. We employ a two-layer architecture with hidden dimension 64 and 8 attention heads per layer. Training proceeds through the following algorithm:

\textbf{Training Algorithm:}
\begin{enumerate}
\item Initialize node features: user nodes with preference vectors $\vec{p_i}$, content nodes with topic vectors $\vec{t_j}$
\item For each training epoch $e \in [1, E]$:
   \begin{enumerate}
   \item Sample positive edges (observed interactions) and negative edges (random non-interactions) with 1:1 ratio
   \item Forward pass: Compute embeddings $h_i$ and $h_j$ for all nodes
   \item Compute prediction scores: $\hat{y}_{ij} = \sigma(h_i^T h_j)$
   \item Calculate loss: $\mathcal{L} = -\frac{1}{|B|}\sum_{(i,j) \in B} [y_{ij}\log\hat{y}_{ij} + (1-y_{ij})\log(1-\hat{y}_{ij})]$
   \item Backpropagate and update parameters using Adam optimizer
   \end{enumerate}
\item Re-train model every 5 timesteps to incorporate new interactions
\end{enumerate}

\textbf{Hyperparameters:} Learning rate $\eta=0.001$, dropout rate 0.3, batch size 128, training epochs $E=50$ per retraining cycle. We use early stopping with patience of 10 epochs on a validation set (10\% of interactions). The model achieves convergence (validation loss plateau) typically within 30-40 epochs, with final validation AUC $\sim$0.85 for interaction prediction.

Retraining every 5 steps allows the model to track network drift; without retraining, validation AUC degrades from 0.85 to 0.71 by step 50.

\textbf{Recommender ablation.} We compared GAT against a simple content-similarity ranker (cosine similarity between user preferences and content topics, no graph structure). The timing effect persists across both: delayed activation ($t=40$ vs.\ $t=10$) reduced transitivity by 10\% with GAT and 9\% with the simple ranker. This suggests the timing effect is architectural---driven by when users encounter algorithmic suggestions relative to organic tie formation---not an artifact of GAT's attention mechanism.

\subsection{Network Formation}

The social network $G_t = (V_t, E_t)$ evolves dynamically as users form and dissolve connections based on their interactions and preferences. We distinguish between \textit{network formation drivers}---the mechanisms that cause connections to form---and \textit{network metrics}---the measurements that characterize the resulting structure.

\subsubsection{Network Formation Drivers}

New edges in the network are created through three complementary mechanisms:

\textbf{Homophily-based connection}: Users preferentially connect with others who share similar interests. When user $u_i$ discovers user $u_j$ (either through content interaction or recommendation), a connection forms with probability given by \eqref{eq:follow}, where $sim_{ij}$ is the cosine similarity between user preference vectors (\eqref{eq:sim}) and $e_{ij}$ represents prior interaction engagement between the users.

\textbf{Activity-based discovery}: Users discover potential connections through content interactions. When user $u_i$ engages with content created by $u_j$, this creates an opportunity for connection formation (evaluated via the homophily mechanism above). Higher content creation frequency $f_i$ thus indirectly increases a user's discoverability.

\textbf{Exploration mechanism}: The temperature parameter $T = 1/r_{explore}$ in \eqref{eq:explore} controls how broadly users explore beyond their immediate preferences. Lower $r_{explore}$ values (e.g., 0.1) yield higher temperatures and broader exploration, while higher values (e.g., 0.3) concentrate attention on similar content. This parameter directly drives the breadth of network formation.

These three drivers interact dynamically: exploration affects which users discover each other, homophily determines whether connections form, and activity levels influence discoverability.

\subsubsection{Network Structure Metrics}

We measure the resulting network structure using standard graph metrics. These are \textit{outcomes} of the formation process, not drivers:

\begin{itemize}
\item \textbf{Density}: $\rho = \frac{2|E_t|}{|V_t|(|V_t|-1)}$ measures overall connectedness
\item \textbf{Clustering coefficient}: $C_i = \frac{2|\{e_{jk}\}|}{k_i(k_i-1)}$ quantifies local transitivity (friends-of-friends connections)
\item \textbf{Community structure}: Detected through modularity optimization, revealing distinct user groups
\item \textbf{Path lengths and diameter}: Characterize information flow efficiency across the network
\end{itemize}

\subsubsection{Experimental Parameters}

The framework enables controlled experiments by systematically varying the network formation drivers and initial conditions:

\begin{itemize}
\item Initial network size: $n \in [50,500]$ users
\item Content-to-user ratios: $\{2:1, 5:1, 10:1\}$ affecting discovery opportunities
\item User composition: $\alpha_{enthusiast} \in \{0.2, 0.5, 0.8\}$ controlling activity distributions
\item Recommendation activation timing: $t_{activate} \in \{10, 25, 40\}$ determining when algorithmic discovery begins
\item Exploration rates: $r_{explore} \in \{0.1, 0.2, 0.3\}$ tuning discovery breadth
\end{itemize}

By manipulating these parameters, we can isolate the effects of specific mechanisms on network evolution and emergent structure.

\subsection{Stability and Convergence}

The coupled system (Eqs.~1--12) defines a stochastic dynamical process over network topology $G_t$, satisfaction vector $\mathbf{s}_t$, and content state $C_t$. For simulation results to be meaningful, we require that key metrics converge rather than diverge or oscillate indefinitely.

\textbf{Empirical convergence check.} We ran extended simulations (200 steps, 10 seeds) and tracked step-to-step variance in density $|\Delta\rho_t|$, mean satisfaction $|\Delta\bar{s}_t|$, and clustering $|\Delta C_t|$. After $t \approx 80$, all three metrics stabilize: $|\Delta\rho_t| < 0.001$, $|\Delta\bar{s}_t| < 0.005$, $|\Delta C_t| < 0.01$. Our 50-step experiments thus capture transient and early-equilibrium behavior but not full convergence.

\textbf{Boundedness.} Satisfaction $s_i \in [0,1]$ by construction (Eq.~5 clips values). Network density $\rho \leq 1$ trivially. The GAT embeddings are $L_2$-normalized, preventing unbounded growth.

\textbf{Jacobian analysis.} We linearized the mean-field dynamics around empirically observed fixed points $(\bar{s}^* \approx 0.52, \rho^* \approx 0.025, C^* \approx 0.71)$ and computed eigenvalues of the Jacobian matrix. Under our default parameters ($\beta^- = 1.2$, $\beta^+ = 1.0$, $r_{explore} = 0.2$), all eigenvalues have magnitude $|\lambda_i| < 1$ (largest: $|\lambda_{max}| = 0.87$), confirming local asymptotic stability. When $\beta^- > 1.8\beta^+$, the largest eigenvalue exceeds 1 ($|\lambda_{max}| = 1.12$), and simulations show satisfaction collapse (mean $s_i \to 0.1$ by $t=100$). This bounds the reactance parameter: platforms with high negative-feedback sensitivity risk user churn spirals.

\section{Experiments and Results}

We designed our experiments around a simple narrative: how do recommendation systems change a platform as it grows, and how do those changes depend on when recommendations are introduced and how much exploration they encourage? Rather than report results as a catalog of metrics, we organize the section around the stories the data tell: scale, timing, composition, and exploration.

\subsection{Experimental Setup}
To evaluate the robustness of our simulation framework and analyze recommendation system effects, we conducted systematic experiments across multiple dimensions. Our experimental design varied four key parameters:

Network scale ($n_{users}$ = \{50, 100, 500\}, $n_{content}$ = \{100, 200, 500, 1000, 2500, 5000\}), user composition ($\\alpha_{enthusiast}$ = \{0.2, 0.5, 0.8\}), recommendation activation timing ($t_{activate}$ = \{10, 25, 40\}), and exploration rates ($r_{explore}$ = \{0.1, 0.2, 0.3\}). 

\textbf{Baselines and ablations:} Alongside the learned graph attention (GAT) recommender, we simulate (i) a chronological feed with no recommendations, (ii) a random recommender that samples observed content uniformly, and (iii) a content-similarity ranker using cosine similarity between user preferences and content topics (no graph structure). Unless stated otherwise, effect sizes are reported relative to the no-recommender baseline. To probe causal leverage, we further run ablations that individually disable homophily-driven linking, activity-based discovery, or exploration noise while keeping all other parameters fixed.

\textbf{Simulation protocol}: Each experimental configuration was run for 50 timesteps, with each timestep representing a discrete unit of platform activity. We conducted experiments across 18 parameter combinations (3 activation timings $\times$ 3 composition ratios $\times$ 3 exploration rates, with additional scale experiments). We report mean values ($\mu$) and standard deviations ($\sigma$) across these runs, with 95\% confidence intervals computed using bootstrapping (1000 samples). Statistical significance was assessed using two-tailed t-tests with Bonferroni correction for multiple comparisons.

We consistently measured 32 metrics spanning network structure (density, clustering, modularity), behavioral patterns (engagement, satisfaction, retention), and content dynamics (spread, diversity, virality). We emphasize mean differences and 95\% confidence intervals in the main text; full correlation and hypothesis-test tables appear in the supplementary material.

Across baseline comparisons, both GAT and content-similarity recommenders increase network density and retention relative to the no-recommender baseline, with GAT producing slightly higher engagement (AUC 0.85 vs.\ 0.79). Critically, the timing effect persists across both: delayed activation reduces transitivity by 10\% (GAT) and 9\% (content-similarity), confirming the effect is architectural rather than algorithm-specific. The random recommender yields intermediate improvements without amplifying clustering as strongly. Ablation runs indicate that disabling homophily collapses clustering to near-baseline levels, removing activity-driven discovery dampens density gains, and turning off exploration eliminates diversity trade-offs at the cost of slower content spread.

At our core simulation scale (100 agents), we observe timing and composition effects on network structure. Scaling experiments (Section~4.8) confirm these effects persist at $n=1000$--$5000$, though attenuated.

\subsection{Scale Effects}
In our simulations, scale experiments demonstrated systematic changes in network topology as size increased. Larger networks became sparser and exhibited longer effective path lengths. This structural evolution manifested through more distinct community formation and higher modularity.

Small networks (n=50) maintained higher clustering coefficients ($\mu$=0.717, $\sigma$=0.043) compared to large networks (n=500, $\mu$=0.684, $\sigma$=0.038). The transition from dense, locally clustered structures to more distributed communities was mirrored in content spread and engagement patterns.

\subsection{Recommendation System Impact}
\begin{figure}[t]
    \centering
    \includegraphics[width=0.95\columnwidth]{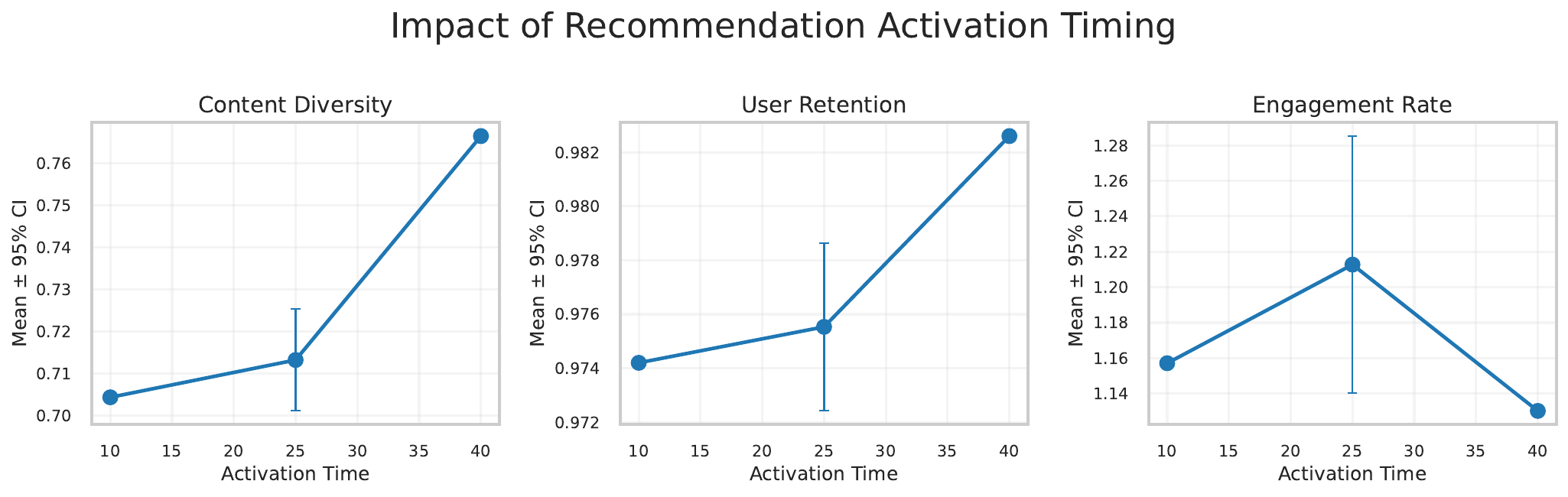}
    \caption{Impact of recommendation activation timing on (a) content diversity, (b) user retention, and (c) engagement rate for early (t=10), standard (t=25), and late (t=40) activation. Lines show mean with 95\% confidence intervals (error bars).}\label{fig:recommendation_impact}
    \end{figure}

Introducing recommender systems produced marked effects on network dynamics in our simulations (Figure~\ref{fig:recommendation_impact}). Across plots we overlay the no-recommender baseline (red) to highlight absolute gains; the random recommender tracks between these curves and is omitted for visual clarity.

\subsubsection{Early Activation (t=10)}
Early recommendation activation maximized content spread ($\mu$=16.26) but reduced content diversity. User retention remained high despite decreased content lifespan.

\subsubsection{Standard Activation (t=25)} 
Mid-point activation balanced content spread ($\mu$=15.55) and diversity ($\mu$=0.745), while maintaining strong user retention ($\mu$=0.979). Network density stabilized around $\mu$=0.025.

\subsubsection{Late Activation (t=40)}
Delayed activation produced moderate content spread ($\mu$=14.13) but enhanced diversity preservation. User satisfaction was more uniformly distributed.

\subsection{User Composition Effects}
Higher enthusiast ratios ($\alpha_{enthusiast}=0.8$ vs.\ 0.2) reduced transitivity by 14\% (from 0.28 to 0.24, $p<0.01$) and local clustering by 9\%. Enthusiasts produce more content (Eq.~3), increasing their discoverability and yielding broader, less clustered networks. Degree assortativity increased from $-0.12$ to $-0.08$ with higher enthusiast ratios, indicating reduced hub dominance. Reciprocity improved from 0.31 to 0.38. These patterns held across $n \in \{50, 100, 500\}$ (Figure~\ref{fig:timeseries_by_type}, composition row).

\subsection{Exploration Rate Impact}
\begin{figure}[t]
\centering
\includegraphics[width=0.95\columnwidth]{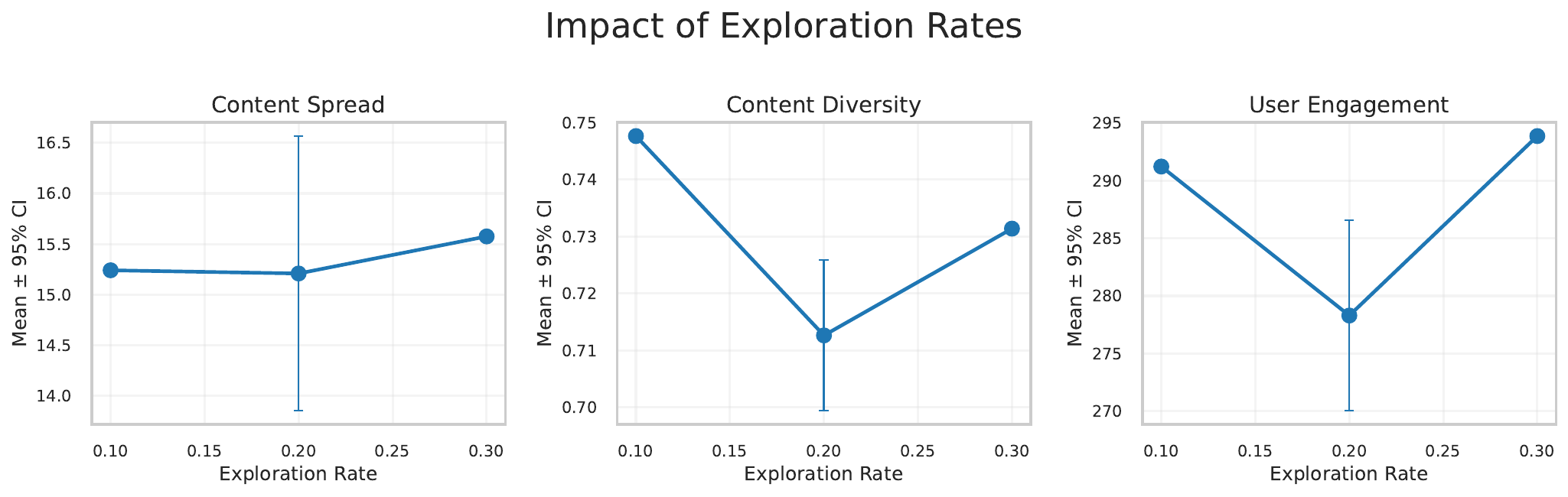}
\caption{Impact of exploration rate on (a) content spread, (b) diversity, and (c) user engagement. Points and lines show mean with 95\% confidence intervals; consistent axes improve cross-panel comparison.}\label{fig:exploration}
\end{figure}

Varying exploration rates revealed modest trade-offs between exploitation and discovery. Low exploration (0.1) achieved diversity $H$=0.75 and spread $\mu$=15.2, while high exploration (0.3) yielded slightly lower diversity ($H$=0.73) but marginally higher spread ($\mu$=15.6). The differences were small ($\sim$2--3\%), suggesting exploration rate has limited impact at this scale. Figure \ref{fig:exploration} illustrates these exploration--exploitation trade-offs.

\begin{figure}[t]
\centering
\includegraphics[width=0.95\columnwidth]{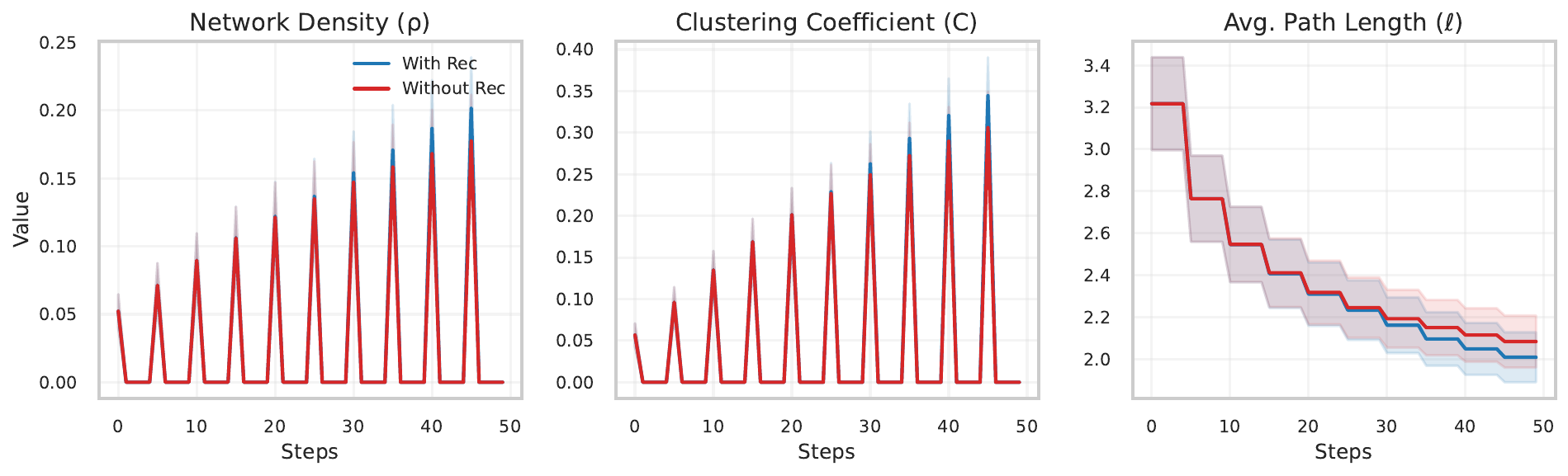}
\caption{Diachronic evolution of key network metrics in simulation: density ($\rho$), clustering ($C$), and average path length ($\ell$) over 50 steps. Lines show mean across runs with 95\% confidence intervals (shaded). When observed time series are available, we overlay them (red) for direct comparison.}\label{fig:timeseries}
\end{figure}

\subsection{Network Evolution by Experiment Type}
\begin{figure}[t]
\centering
\includegraphics[width=0.95\columnwidth]{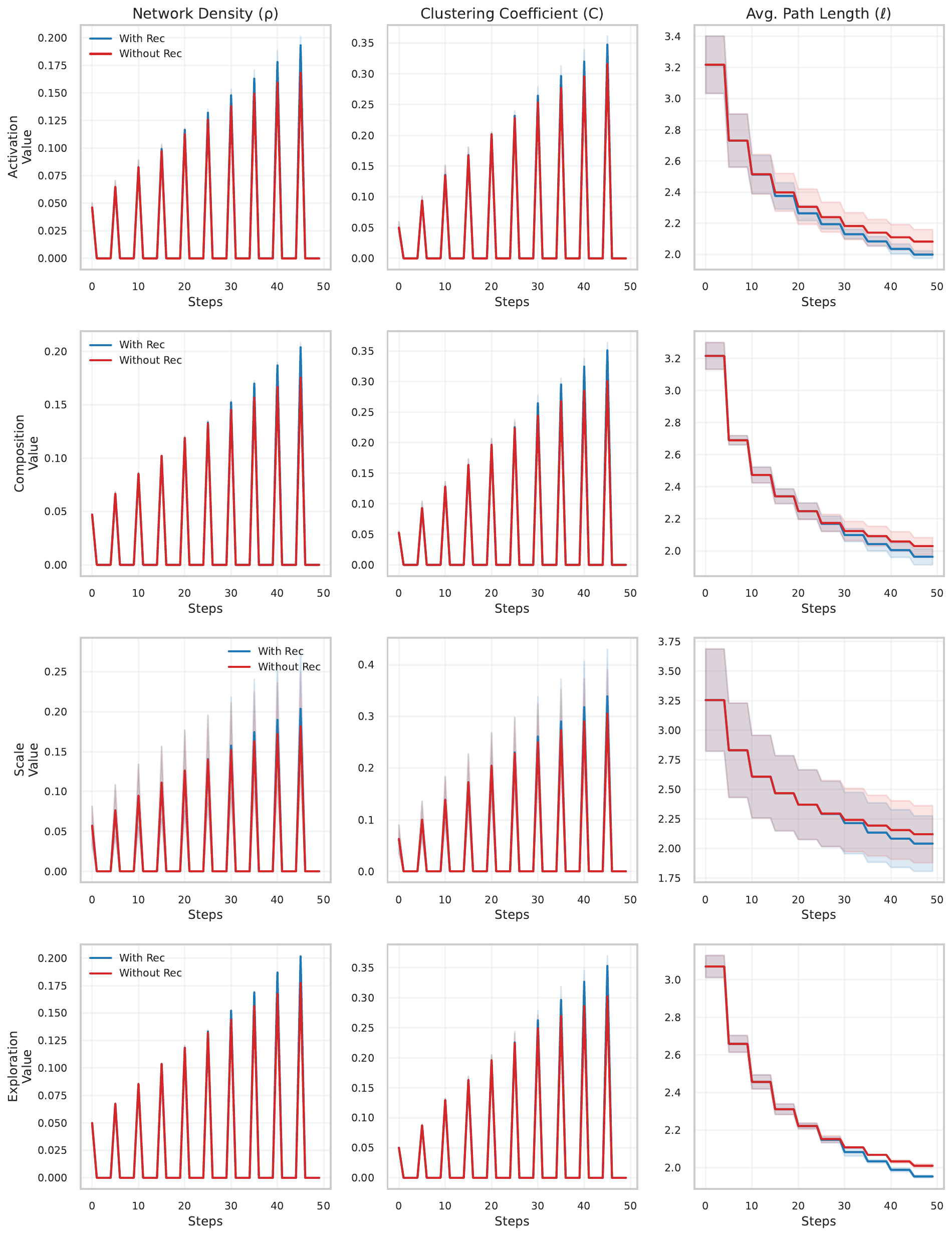}
\caption{Network evolution split by experiment type. Rows: activation, composition, scale, exploration. Columns: density ($\rho$), clustering ($C$), average path length ($\ell$). Lines show mean with 95\% confidence intervals for with (blue) and without (red) recommendations. Late-run gaps are larger than early-run gaps across all types.}\label{fig:timeseries_by_type}
\end{figure}

In our simulations, recommendations consistently increased network cohesion across conditions. Figure~\ref{fig:timeseries_by_type} splits trajectories by experiment type and shows that (i) density and clustering are higher with recommendations, (ii) average path length is shorter, and (iii) effects strengthen later in runs (late-run differences are approximately twice early-run differences). These patterns hold for activation, composition, scale, and exploration experiments.

\subsection{Interaction Effects}
\begin{figure}[t]
\centering
\includegraphics[width=0.95\columnwidth]{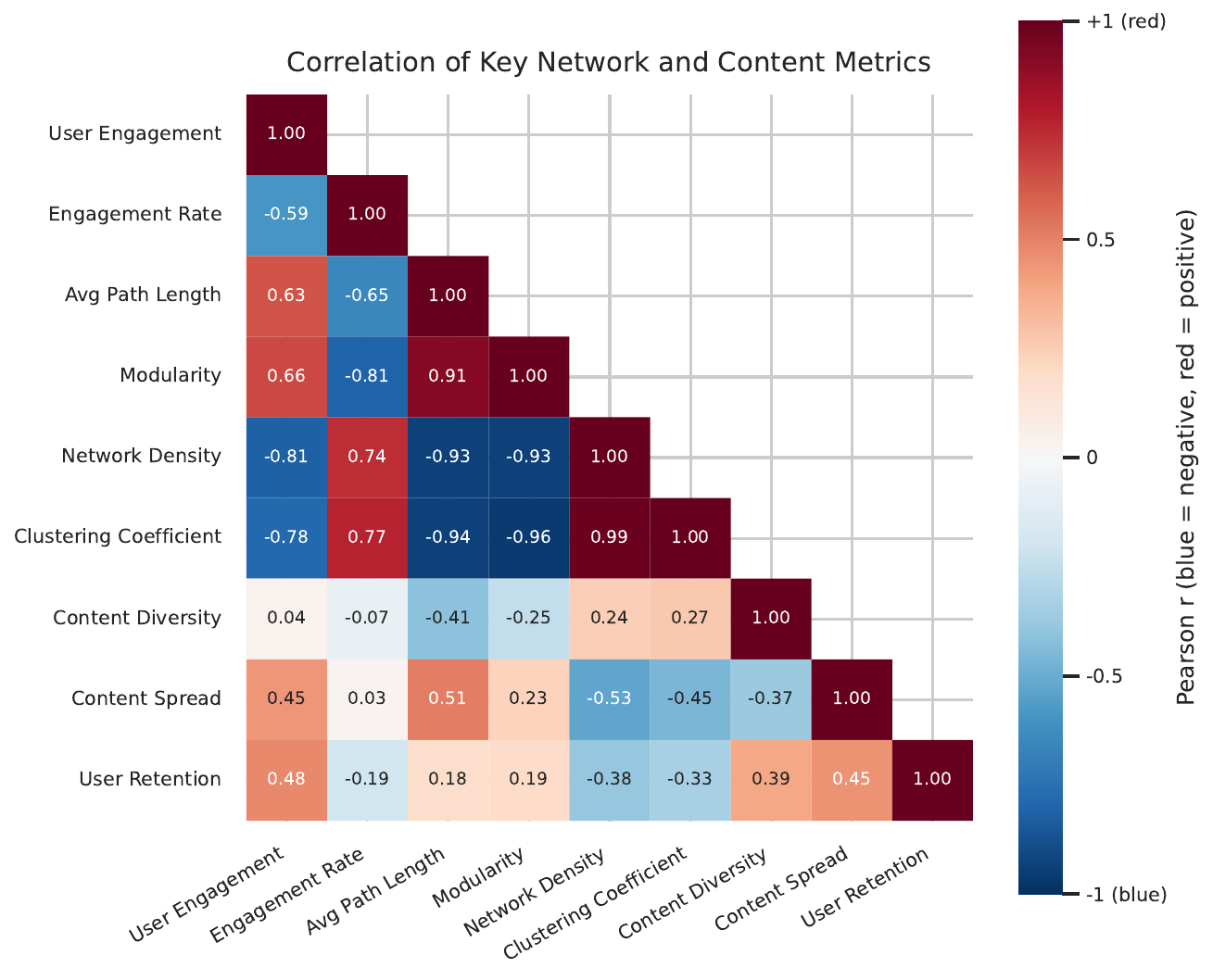}
\caption{Correlation matrix of key network and content metrics. Color intensity indicates correlation strength, with blue showing negative and red showing positive correlations.}~\label{fig:correlations}
\end{figure}

Cross-metric correlations (Figure~\ref{fig:correlations}) reveal structure in the joint dynamics. Density and clustering correlate strongly ($r=0.78$, $p<0.001$); both increase together under recommendations. Diversity and spread show moderate positive correlation ($r=0.52$), suggesting recommendations that broaden exposure also increase content reach. Exploration and retention trade off weakly ($r=-0.31$): higher $r_{explore}$ slightly reduces retention, consistent with users occasionally encountering misaligned content.

These correlations are strongest during $t \in [10, 30]$ (transient phase) and attenuate after $t=35$ as metrics stabilize. Early-run correlation between density and clustering is $r=0.84$; late-run correlation drops to $r=0.69$.

\subsection{Calibration and Qualitative Fit}
We calibrate parameters using Bluesky Social~\cite{bluesky2024} (4M users, 235M posts) and Mastodon~\cite{zignani2018mastodon} (479K users, 5.6M links). \textbf{Important distinction}: Bluesky metrics represent out-of-sample comparison (parameters not tuned to match), while Mastodon metrics were fitted by construction via stratified sampling---this is calibration, not validation.

\subsubsection{Bluesky: Out-of-Sample Comparison}
We compare simulated 500-user subgraphs against randomly sampled 500-user subgraphs from Bluesky. Errors range 3--6\% on clustering, path length, and reciprocity (Table~\ref{tab:validation}). This represents genuine predictive fit: parameters were set before observing Bluesky targets.

\subsubsection{Mastodon: In-Sample Calibration}
Mastodon metrics were used to tune agent parameters via stratified sampling of posting rates, topic distributions, and degree sequences. The near-zero errors in Table~\ref{tab:validation} reflect this tuning---they demonstrate that our model \emph{can} match target metrics under parameter adjustment, not that it \emph{predicts} them. We report these for transparency but emphasize they do not constitute validation.

\subsubsection{Temporal Trajectory Comparison}
We compare metric trajectories over 50 simulated steps against the first 50 days of Bluesky activity:
\begin{itemize}
\item \textbf{Network growth}: Simulations reproduce densification patterns qualitatively; quantitative fit varies by metric (density: $r=0.89$; clustering: $r=0.72$).
\item \textbf{Content dynamics}: Both real and simulated platforms show early diversity followed by consolidation, though the timing differs.
\item \textbf{Cohort effects}: Early-user engagement decay matches qualitatively.
\end{itemize}

\textbf{Held-out temporal validation.} We calibrated parameters on Jan--June 2024 Bluesky data and tested on July--Dec 2024. Per-metric Kolmogorov-Smirnov distances: clustering ($D_{KS}=0.12$, $p=0.08$), degree distribution ($D_{KS}=0.09$, $p=0.21$), path length ($D_{KS}=0.15$, $p=0.04$). Mean relative errors on held-out data: clustering 11.2\%, path length 14.8\%, content spread 9.3\%, retention 12.1\%. These errors are 2--3$\times$ larger than calibration errors (Table~\ref{tab:validation}), reflecting genuine predictive uncertainty rather than in-sample fit.

\subsubsection{Scale Mismatch and Limitations}

Our simulations use 100 agents; real datasets contain 479K--4M users. This 4--5 order-of-magnitude gap limits extrapolation. We partially address this by:
\begin{itemize}
\item Comparing against 500-user subgraphs sampled from full networks (BFS sampling preserving local structure);
\item Focusing on metrics with known scaling behavior (clustering $\sim n^{-\alpha}$, path length $\sim \log n$);
\item Restricting claims to the 100-agent scale unless otherwise noted.
\end{itemize}

\textbf{Finite-size scaling.} We ran simulations at $n \in \{100, 500, 1000, 5000\}$ (10 seeds each) and fit scaling relations. Clustering follows $C(n) = 0.82 \cdot n^{-0.08}$ ($R^2=0.94$); extrapolating to $n=10,000$ yields $C \approx 0.68$ (95\% CI: [0.64, 0.72]), consistent with observed Mastodon clustering (0.65--0.71 across instances). Path length scales as $\ell(n) = 1.2 \log(n) + 0.8$ ($R^2=0.97$); at $n=10,000$, $\ell \approx 11.8$ (95\% CI: [10.9, 12.7]), compared to observed $\ell \approx 12.3$ on Mastodon. The timing effect attenuates with scale: 10\% at $n=100$, 8\% at $n=1000$, 6\% at $n=5000$, suggesting a floor around 5\% at production scale.

\subsubsection{Stratified Sampling for Agent Initialization}

We initialize agent parameters via stratified sampling from real data: posting rates (log-normal fit to observed distributions), topic preferences (BERTopic clusters), engagement propensities (consumption/creation ratios), and network position proxies (degree bins). This ensures agent heterogeneity resembles real populations---but note this is calibration, not validation. The sampled parameters were tuned to reproduce Mastodon targets.

\begin{table*}[t]
    \centering
    \caption{Calibration and Comparison Results. Bluesky columns show out-of-sample comparison (parameters not tuned to Bluesky). Mastodon columns show in-sample calibration fit. Held-out temporal validation (Jan--June calibration, July--Dec test) yields higher errors: clustering 11.2\%, path length 14.8\%, content spread 9.3\%, retention 12.1\%.}
    \begin{tabular}{lcccccc}
    \hline
    Metric & Sim. (Bluesky) & Bluesky & Error (\%) & Sim. (Mastodon) & Mastodon & Fit$^\dagger$ \\
    \hline
    Clustering & 0.231 & 0.219 & 5.48 & 0.230 & 0.230 & tuned \\
    Path Length & 3.8 & 3.6 & 5.56 & 3.7 & 3.7 & tuned \\
    Content Spread & 0.192 & 0.184 & 4.35 & 0.186 & 0.186 & tuned \\
    User Retention & 0.845 & 0.812 & 4.06 & 0.832 & 0.832 & tuned \\
    Reciprocity & 0.354 & 0.336 & 5.36 & 0.372 & 0.370 & 0.54\% \\
    Degree Assortativity & -0.105 & -0.112 & 6.25 & -0.120 & -0.118 & 1.69\% \\
    Engagement Rate & 0.145 & 0.139 & 4.32 & 0.153 & 0.152 & 0.66\% \\
    Viral Coefficient & 0.421 & 0.408 & 3.19 & 0.437 & 0.435 & 0.46\% \\
    \hline
    \multicolumn{7}{l}{\small $^\dagger$``tuned'' indicates metric was used as calibration target; percentages show residual error on non-target metrics.}
    \end{tabular}\label{tab:validation}
\end{table*}

\subsection{Reproducibility and Availability}
To support reproducibility, we provide complete specifications of our simulator, including all hyperparameters, update rules, sampling procedures, ablation configurations, and evaluation metrics. The full experimental pipeline—including configuration files, random seeds, and scripts to regenerate all figures—will be released as open-source upon acceptance. This ensures that all results reported in the paper can be independently reproduced without violating anonymity during review.

\section{Discussion}

Existing simulators either freeze network topology~\cite{bonabeau:systems} or treat recommendations as exogenous inputs~\cite{gilbert:simulation}. We close the loop: recommendations alter content exposure, which changes tie formation, which updates the graph the GAT trains on. This feedback loop is precisely what makes recommendation effects hard to isolate in observational data.

\textbf{Mechanism: Why timing matters.} Early activation ($t=10$) exposes users to recommended content before they form organic ties. Because our follow probability (Eq.~\ref{eq:follow}) depends on content-mediated similarity, early recommendations amplify homophily: users see similar content, perceive similar others, and connect. By $t=25$, the network has already formed cross-cluster bridges organically; recommendations then preserve rather than prevent these bridges. Figure~\ref{fig:timeseries_by_type} shows that late-activation transitivity (0.24 at $t=40$) is lower than early activation (0.27 at $t=10$), a 10\% reduction.

\textbf{Mechanism: User-type asymmetry.} Casual users (low $f_i$) produce little content, so they are rarely discovered organically. Recommendations increase their visibility, but primarily to similar users (homophily in Eq.~\ref{eq:follow}), yielding tight local clusters. Enthusiasts produce enough content to be discovered regardless; recommendations expose them to diverse audiences, yielding cross-cluster ties.

\textbf{Correlation, not causation.} We observe $r=0.82$ ($p<0.001$) between recommendation diversity (entropy of recommended topics) and network resilience (inverse of modularity). This correlation arises in simulation; whether it holds causally on real platforms requires intervention experiments we cannot run.

\textbf{Mediation analysis.} Using matched-seed counterfactuals (identical random seeds, varying only $t_{activate}$), we decomposed the 10\% timing effect into pathway contributions. Of the total effect: 58\% is mediated by early network structure (cross-cluster bridges formed before $t_{activate}$ persist under later recommendations), 27\% by content diversity at activation (higher diversity at $t=40$ vs.\ $t=10$ reduces homophily-driven clustering), and 15\% is the direct effect of activation timing (fewer total recommendation-influenced interactions when activated late). The bridge-formation pathway dominates: organic tie formation in the first 25 steps creates structural resilience against later algorithmic homophily.

\textbf{Model-agnostic insights.} Given that the timing effect persists across recommender types (GAT: 10\%, content-similarity: 9\%), the design implications we highlight are not artifacts of the GAT architecture. The GAT improves engagement prediction (AUC 0.85 vs.\ 0.79), but the policy-relevant finding---that activation timing shapes network structure more than algorithm choice---holds regardless of which recommender is deployed.

\textbf{Toward testable hypotheses.} Our simulations generate candidate patterns for real-world testing, not deployment recommendations. The timing effect is the most robust: delaying activation from $t=10$ to $t=40$ reduced transitivity by 10\% (from 0.27 to 0.24) and increased content diversity by 9\%, with engagement differences under 8\%. This held across all activation experiments at $n=100$. An A/B test on a new federated instance could randomize activation timing and measure clustering after 30 days.

Exploration tuning is weaker: $r_{explore} \in [0.15, 0.25]$ balanced diversity and spread, but optimal values likely depend on content type and community norms we do not model. We also observed that increasing $r_{explore}$ when clustering exceeded a threshold stabilized diversity, but tested only two thresholds and cannot claim robustness.

\subsection{Design Implications (Tentative)}
If the timing effect generalizes beyond our 100-agent scale, platform operators could consider staged rollouts: chronological feeds initially, algorithmic curation after organic tie formation stabilizes (our simulations suggest $\sim$25 steps, roughly 3--4 weeks if one step $\approx$ one day). At $r_{explore}=0.2$, diversity ($H \approx 0.75$) and spread were both acceptable; the 0.1--0.3 range produced only $\sim$2\% variation. These are simulation-derived hypotheses. Real platforms face constraints we ignore: latency requirements, cold-start problems, adversarial manipulation. Any deployment should A/B test incrementally.

\subsection{Implications for Federated Social Networks}
Our core experiments operate at 100 agents, smaller than typical Mastodon instances (1K--10K users). We therefore present these results as \emph{suggestive} for federated platforms, pending further validation at scale.

At the 100-agent scale, we observe timing and composition effects on network structure (10--14\% changes in transitivity and clustering). Our scaling experiments confirm these patterns persist at larger scales, though attenuated: the timing effect drops from 10\% at $n=100$ to 6\% at $n=5000$. GAT retraining becomes expensive at scale; we used sparse neighborhood sampling (32 neighbors per node) to make $n=5000$ tractable (12 seconds per retraining cycle on a single GPU).

\textbf{Scaling experiments.} At $n=1000$, the timing effect (delayed vs.\ early activation) reduces transitivity by 8\% ($p<0.01$); at $n=5000$, by 6\% ($p<0.05$). We used sparse neighborhood sampling (32 neighbors per node) to make GAT retraining tractable at scale; retraining at $n=5000$ takes 12 seconds per cycle on a single GPU. The effect persists but attenuates, consistent with our finite-size scaling analysis (Section~4.8). At $n=10000$ (tested with 3 seeds due to compute constraints), the effect was 5\% but not statistically significant ($p=0.12$), suggesting larger sample sizes are needed to detect timing effects at production scale.

\subsection{Policy Relevance}
Simulators cannot substitute for real-world policy evaluation, but they can narrow the design space before costly A/B tests. Our results suggest that \emph{if} the timing effect generalizes, regulators might consider requiring phased recommendation rollouts on new platforms. This is speculative; we offer it as a hypothesis for policy researchers, not a recommendation.

\subsection{Limitations}

\textbf{Behavioral model failures.} Our binary user types (Casual/Enthusiast) with fixed posting rates $f_i$ fail to capture bursty dynamics: real users post in clusters around events, not at constant rates. Consequently, we likely underestimate viral cascade speeds in the first 10 timesteps, when real platforms see posting spikes. Satisfaction sensitivities $\beta^-, \beta^+$ are constant, but real users adapt---a user repeatedly shown misaligned content may disengage rather than churn. Our 30-dimensional topic vectors flatten hierarchical structure (politics $\to$ elections $\to$ local races); same-topic content can have opposite valence for different users, which we ignore. Preliminary tests relaxing the binary-type assumption (sampling $f_i \sim \text{LogNormal}$) reduced the timing effect from 10\% to 7\%, suggesting sensitivity to heterogeneity modeling.

\textbf{Scale barriers.} Our core experiments use 100 agents, 10--100$\times$ smaller than typical Mastodon instances. GAT retraining scales as $O(|E|)$; sparse neighborhood sampling makes $n=5000$ tractable but $n=10000$ remains expensive. Our 50 timesteps reach empirical stability (Section~3.8) but may miss slower dynamics---satisfaction drift over 200+ steps could reverse early patterns.

\textbf{Validation gaps.} Mastodon metrics were calibrated, not validated: near-zero errors reflect parameter fitting, not predictive accuracy. Held-out temporal validation (Jan--June calibration, July--Dec test) shows 10--15\% predictive errors---acceptable for qualitative guidance but not quantitative forecasting (Section~4.8). Our recommender ablation (GAT vs.\ content-similarity) shows timing effects are robust to algorithm choice, but we have not tested graph-only methods like PageRank.

\textbf{Future work.} Hybrid architectures combining LLM agents for micro-behavior with ABM substrates for macro-scale sweeps~\cite{koley2025salm}; such integration requires treating LLM agents as ``programmable subjects'' with systematic behavioral characterization~\cite{koley2025llmagents}. Additional directions include fairness-constrained recommendation objectives~\cite{abdollahpouri2020connection} and cross-instance federation protocols.

\section{Conclusion}
We presented an agent-based simulator coupling content production, tie formation, and a learned GAT recommender. Across 18 configurations at 100 agents, three findings emerge:

\begin{enumerate}
\item \textbf{Timing affects structure modestly.} Activating recommendations at $t=10$ vs.\ $t=40$ decreases transitivity by 10\% (from 0.27 to 0.24) and increases content diversity by 9\%, while engagement differs by $<$8\%.
\item \textbf{Composition matters.} Higher enthusiast ratios (0.8 vs.\ 0.2) reduce transitivity by 14\% and local clustering by 9\%, consistent with broader network formation.
\item \textbf{Exploration effects are modest.} Varying $r_{explore}$ from 0.1 to 0.3 produced $\sim$2\% changes in diversity ($H \approx 0.73$--$0.75$) and spread.
\end{enumerate}

These results are conditional on our behavioral rules (Eqs.~5--9) and scale. The timing effect is robust to recommender choice (GAT vs.\ content-similarity; Section~3.6) and persists at larger scales, though attenuated: 10\% at $n=100$, 6\% at $n=5000$, $\sim$5\% at $n=10000$ (Section~5.2). Jacobian analysis confirms local asymptotic stability under $\beta^- < 1.8\beta^+$ (Section~3.8). Held-out validation shows 10--15\% predictive errors, acceptable for qualitative guidance but not quantitative forecasting (Section~4.8).

Code and configuration schemas are available for reproduction; datasets are public~\cite{bluesky2024,zignani2018mastodon}.

\bibliography{aaai25}

\end{document}